\documentclass[12pt]{aastex}

\begin{document}

\title{Finding extraterrestrial life using ground-based high-dispersion spectroscopy} 
\author{I.A.G. Snellen$^1$, R.J. de Kok$^2$, R. le Poole$^1$, M. Brogi$^1$, \&  J. Birkby$^1$ 
\\ 
$^1$Leiden Observatory, Leiden University, Postbus 9513, 2300 RA Leiden, The Netherlands\\
$^2$SRON, Sorbonnelaan 2, 3584 CA Utrecht, The Netherlands}
\begin{abstract}
Exoplanet observations promise one day to unveil the presence of extraterrestrial life. Atmospheric compounds in strong chemical disequilibrium would point to large-scale biological activity just as oxygen and methane do in the Earth's atmosphere. The cancellation of both the Terrestrial Planet Finder and Darwin missions means that it is unlikely that a dedicated space telescope to search for biomarker gases in exoplanet atmospheres will be launched within the next 25 years. Here we show that ground-based telescopes provide a strong alternative for finding biomarkers in exoplanet atmospheres through transit observations. Recent results on hot Jupiters show the enormous potential of high-dispersion spectroscopy to separate the extraterrestrial and telluric signals making use of the Doppler shift of the planet. The transmission signal of oxygen from an Earth-twin orbiting a small red dwarf star is only a factor 3 smaller than that of carbon monoxide recently detected in the hot Jupiter $\tau$ Bo\"otis b, albeit such a star will be orders of magnitude fainter. We show that if Earth-like planets are common, the planned extremely large telescopes can detect oxygen within a few dozen transits. Ultimately, large arrays of dedicated flux collector telescopes equipped with high-dispersion spectrographs can provide the large collecting area needed to perform a statistical study of life-bearing planets in the solar neighborhood.
\end{abstract}


\section{Introduction}

We investigated the strength of biomarker signatures (Lovelock 1965; Lovelock \& Hitchcock 1967) in twin-Earth atmospheres that can be targeted with ground-based high-dispersion spectroscopy. At wavelengths $>$5 $\mu$m the sky background is so high that its brightness per square arcsecond compares to that of a sun-like star of V=0 magnitude. This practically makes ground-based planet characterization impossible in this wavelength regime, including the ozone feature at 9.6 $\mu$m. In addition, telluric absorption bands block significant parts in the near-infrared, except for the well-known spectral windows.  Since by definition telluric absorption is identical to absorption in a twin-Earth atmosphere, only weak features can be probed, away from the heart of the main molecular bands. It is therefore unlikely that ground-based observations targeting H$_2$O, CO$_2$ or CH$_4$ in Earth-like planets could compete with those obtained from space (e.g. Kaltenegger \& Traub 2009; Charbonneau \& Deming 2007) - even if the latter are conducted with significantly smaller telescopes. 

Fortunately, the case is very different for molecular oxygen, which exhibits two strong absorption bands at 0.76 $\mu$m and 1.26 $\mu$m, both relatively isolated from other sources of absorption in the Earth spectrum. Importantly, at high spectral dispersion the bands resolve into sets of strong well-separated lines, with near-100\% transmission in between. This is ideal for high-dispersion exoplanet spectroscopy. Firstly, the extraterrestrial signals can be observed in between the telluric absorption lines, making use of the radial velocity of the exoplanet and the barycentric velocity of the Earth. Secondly, the concentration of the oxygen absorption in strong well-separated lines maximizes the signal at high spectral resolution. Since the star/planet contrasts at these wavelengths are very high in the range of 10$^{7-10}$, transmission spectroscopy is significantly favourable over dayside spectroscopy.

\section{Simulation of oxygen in transmission in a twin-Earth atmosphere}

\begin{figure}
\begin{center}
\includegraphics[width=0.45\textwidth]{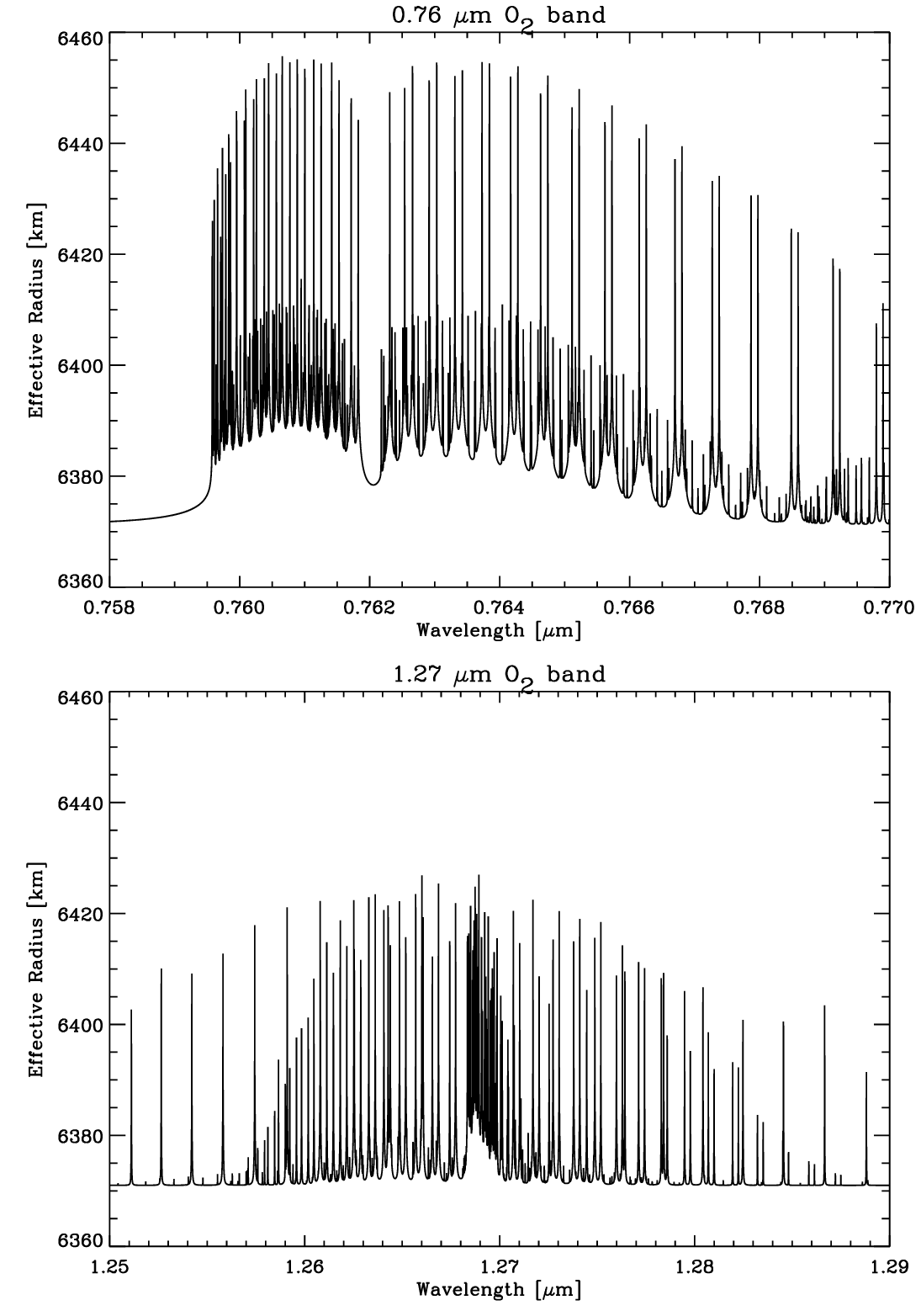}
\end{center}

\caption{\label{o2spec} (top panel) Simulated transmission spectrum of the Earth covering the 0.76 $\mu$m O$_2$ A-band, showing the effective planet radius as function of wavelength at a spectral resolution of R=100,000. Clouds (not included here) typically contribute to an altitude of 5-10 km and only affect the base of the spectrum, not the strength of the narrow O$_2$ lines. (bottom panel) The same simulation covering the 1.27 $\mu$m O$_2$ band.}
\end{figure}
\begin{figure*}
\begin{center}
\includegraphics[width=1.0\textwidth]{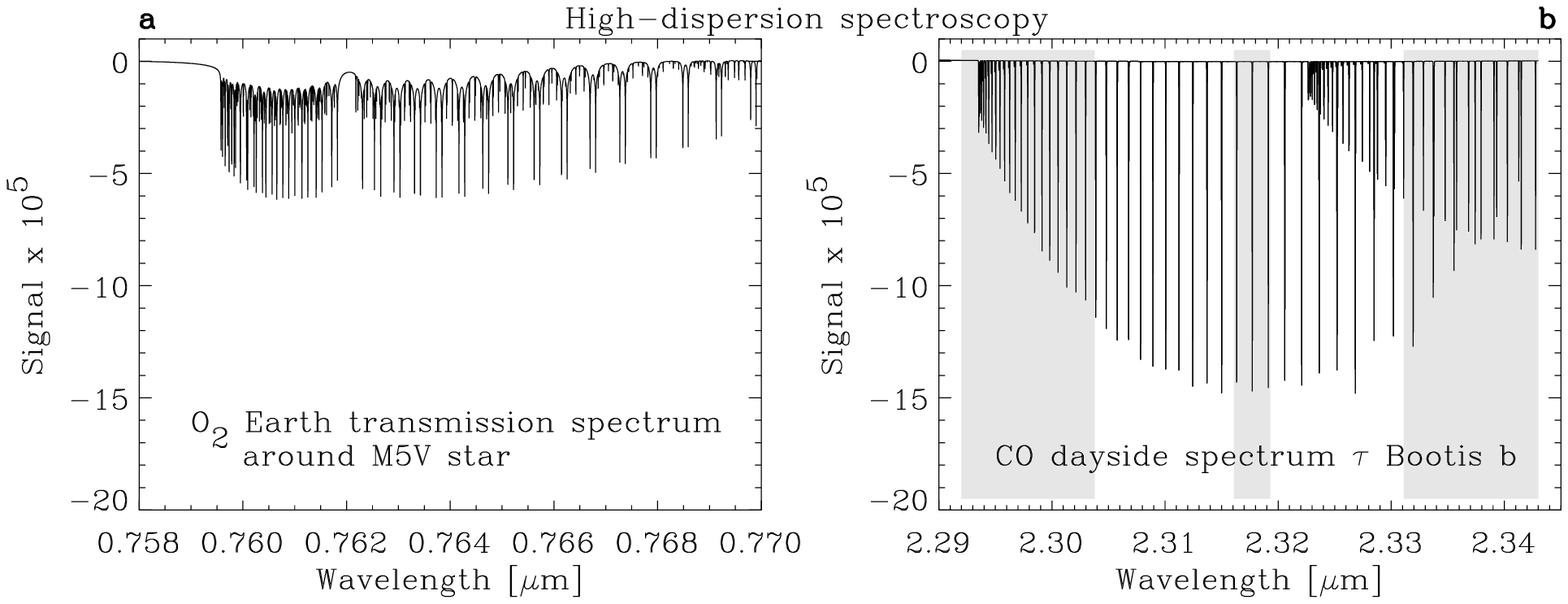}
\end{center}
\caption{\label{comp} The left panel shows the simulated O$_2$ transmission signal for an Earth-twin transiting an M5 dwarf star. The wavelength range is centered on the oxygen 0.76 $\mu$m A-band. The transit signal of the bulk of the planet is removed. The right panel shows the model spectrum that fitted best the CO detection in the dayside spectrum of the non-transiting hot Jupiter $\tau$ Bo\"otis b (Brogi et al. 2012). The white areas are those wavelength ranges covered by the observations with the Very Large Telescope. It shows that the predicted O$_2$ transmission signal from an Earth-twin in the habitable zone of an M5 red dwarf star is only a factor 3 lower than the CO signal detected by Brogi et al. (2012).}
\end{figure*}

We calculated the oxygen transmission in the two bands at a spectral resolution of R=100,000 through a spherical atmosphere with 50 layers between 1 and 1x10$^{-9}$ bar (roughly 0-140 km), integrating over the entire limb of the Earth. A constant oxygen volume-mixing ratio of 20.95\% was assumed up to the top of the mesosphere at 85 km altitude. The gas opacity was calculated line-by-line using a Voigt profile. Opacity data for O$_2$ were obtained from the HITRAN 2008 database (Rothman et al. 2009). The resulting effective planet radius as function of wavelength for the 0.76 $\mu$m A-band is shown in Figure 1. At this resolution, the transmission spectrum is dominated by approximately 50 strong lines which each increase the effective radius of the planet by 60-80 km, up to an order of magnitude higher than the Earth atmospheric scale height of H=8 km. Convolved to low spectral resolution, the oxygen transmission signal decreases to 10-20 km - similar to that calculated in earlier studies (Tinetti et al. 2006; Ehrenreich et al. 2006; Kaltenegger et al. 2009; Rauer et al. 2011; Palle et al. 2011). Importantly, at low spectral resolution the oxygen transmission signal is further decreased by clouds, which virtually play no role at high dispersion since located at an altitude of 5-10 km they only influence the base of the transmission spectrum. The 1.26 $\mu$m oxygen band has a similar number of lines, but is overall ~30\% weaker (Figure 1). 

A planet transit signal is inversely proportional to the square of the radius of the host star, making transmission spectroscopy order(s) of magnitude more powerful for planets transiting late M-dwarfs than for those transiting sun-like stars. The O$_2$ transmission spectrum of an Earth-twin in front of an M5 red dwarf star is shown in the left panel of Figure 2. The individual lines are observable at a level of 3-5x10$^{-5}$ with respect to the stellar flux. We compare this to the best-fit model of the carbon monoxide signal already detected (Brogi et al. 2012) in the dayside spectrum of $\tau$ Bo\"otis b, as shown in the right panel at the same scale. Remarkably, the O$_2$ lines are only a factor of 2-3 times weaker than those from carbon monoxide observed with current-day instrumentation, showing the exciting possibility that oxygen in the atmosphere of an Earth-twin can be detected in this way. 

\section{The brightest transiting twin-Earth systems}

The challenge comes from the expected faintness of the nearest late M-dwarf with a transiting Earth-twin, requiring significantly larger telescope collecting areas than currently used. We have estimated the expected brightness of the most nearby transiting Earth-twin systems considering three types of host stars (see Table 1); sun-like stars (spectral type G0-G5), early M-dwarfs (M0-M2), and late M-dwarfs (M4-M6). The local mass function as determined by the RECONS team\footnote{http://www.recons.org/mf.2009.0.html} provides the necessary statistics of the number of stars as function of distance (and apparent magnitude). The stellar effective temperature and radius (Cox 2000) determine the orbital radius for a planet in the habitable zone and subsequently its transit probability. For each stellar type and planet frequency this is used to calculate the 1$\sigma$ (67\%) probability range for the I-band magnitude of the brightest transiting Earth-twin system for a given $\eta_e$. In the most favourable case that all stars have an Earth-like planet in their habitable zone ($\eta_e$ =100\%), the brightest transiting system is expected between I=4.4 and 6.1 magnitudes for solar type stars, and between I=10.0 and 11.8 for late M-dwarfs. For different values of $\eta_e$ the ranges shift towards fainter magnitudes as $\frac{5}{3}Log_{10} \eta_e$. If only 10\% of the stars have such planets, then the nearest systems are expected to be fainter by 1.7 magnitudes. Note that recently, a first estimate of the frequency of Earth-mass planets in the habitable zone of M-dwarfs was obtained by Bonfils et al. (2012), with 0.28 $<\eta_e<$ 0.95.

In principle, the signal-to-noise ratios (SNR) per transit for the brightest systems should be similar for sun-like host-stars and late M-dwarfs (if $\eta_e$ is independent of spectral type), because the effects of transit depth, duration, and expected magnitude of the host star cancel out. However, it will be significantly more difficult to suppress systematic effects down to the 10$^{-6}$ level to reach the photon noise required for systems of sun-like stars, compared to a few times 10$^{-5}$ for M5 dwarfs - a level that is already almost being reached today (Snellen et al. 2010; Brogi et al. 2012). Furthermore, since observations of multiple transits will need to be combined, the planet orbital period is an important factor. An Earth-twin around a sun-like star would transit only once every year, with at best only part of the 13-hour transit observable from a particular location on Earth. The orbital period of a planet in the habitable zone of an M5 dwarf is 11.8 days, making it all-in-all more than an order of magnitude faster to reach a certain detection level by combining multiple transits. 

The effective transmission signal from the oxygen lines in the 1.27 $\mu$m band is about 60\% of that from the 0.76 $\mu$m band (taking into account a cloud deck at 10 km altitude). While this strongly favors targeting the optical band for early type stars, in case of mid-M dwarfs this decrease in signal is compensated for by the increase of flux received at near-infrared wavelengths. However, the preference for CCD detector technology still drives this balance to the optical, also for mid-M dwarfs. Only for the coolest stars (M7 and beyond) it may be beneficial to target the 1.27 $\mu$m band.

\begin{table*}
\begin{tabular}{ccccccccccr}\hline
Stellar & R$_*$&M$_*$&a$_{HZ}$&Prob&P$_{HZ}$&Dur.&I ($\eta_e$=1)& Line &SNR&Time\\
 type    & [R$_{\rm{sun}}$] & [M$_{\rm{sun}}$] & [au] & [\%] & [days] & [hrs] & [mag] & Contrast &$\sigma$& (yrs) \\ \hline
G0-G5&1.00&1.00&1.000&0.47&365.3&13&4.4 - 6.1&2$\times 10^{-6}$&1.1-2.5&80-400\\
M0-M2&0.49&0.49&0.203&1.12&47.7&4.1&7.3 - 9.1&8$\times 10^{-6}$&0.7-1.5&20-90\\
M4-M6&0.19&0.19&0.058&1.52&11.8&1.4&10.0-11.8&5$\times 10^{-5}$&0.7-1.7&4-20\\ \hline
\end{tabular}
\caption{Properties of hypothetical stellar systems with transiting Earth-twins in their habitable zone. Columns 1 to 7 give the range in stellar types considered, the stellar radius and mass, the orbital distance of a planet in the habitable zone, the transit probability, the orbital period, and the transit duration. Column 8 gives 1$\sigma$ range in I-band magnitude expected for the most nearby and brightest transiting Earth-twin for a planet frequency of $\eta_e$=100\%. Column 9  gives the contrast ratio for the strongest lines in the O$_2$ A-band with respect to the stellar continuum. Column 10  gives the theoretical signal-to-noise achieved with a 39m ELT per transit, taking the combined signal from the 50 lines, and column 11 the total elapsed time before enough transits can be observed from one location on Earth such that an SNR of 5 is reached. }
\end{table*}

\section{Extremely Large Telescope simulations}

\begin{figure}
\begin{center}
\includegraphics[width=0.45\textwidth]{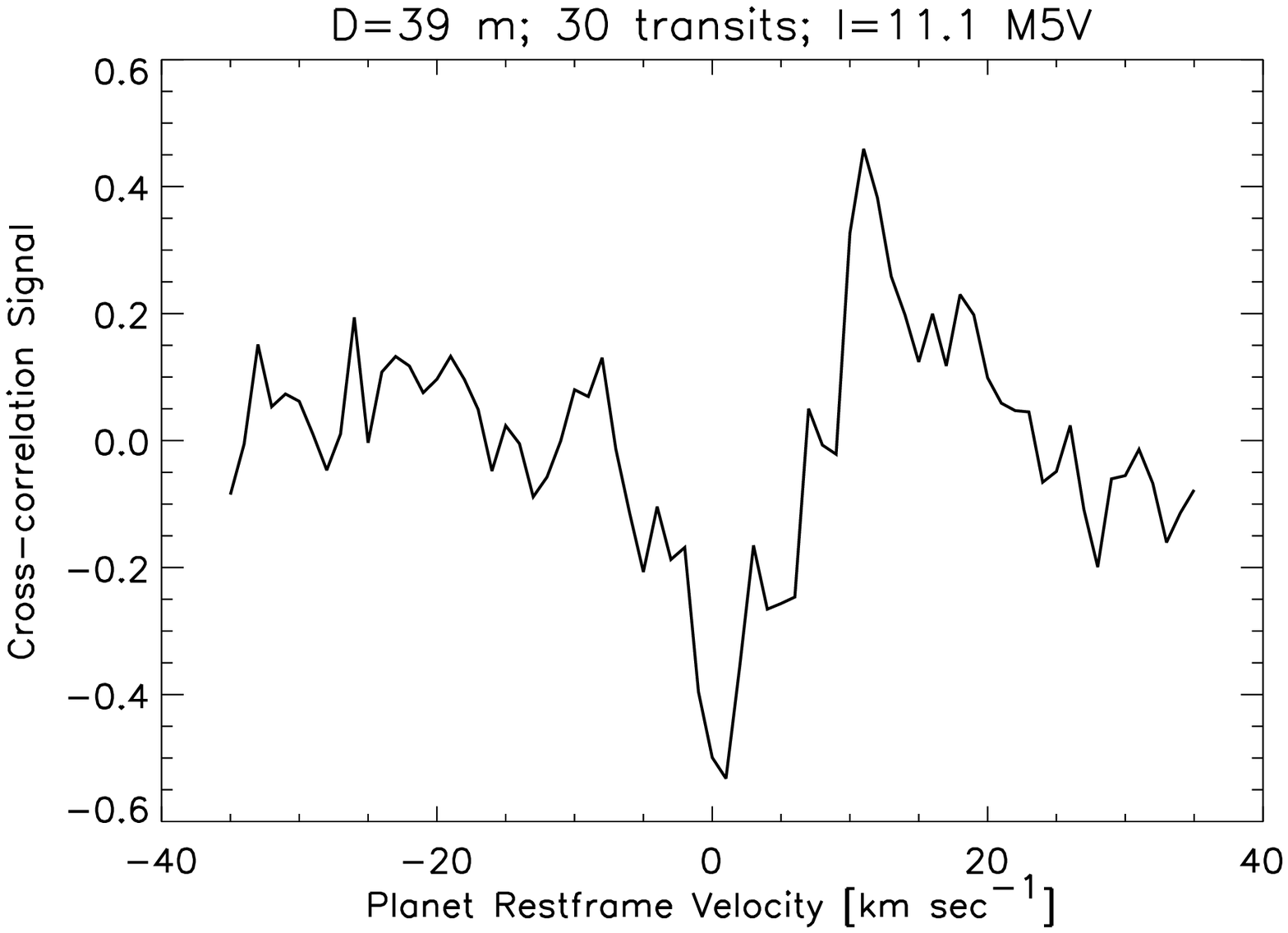}
\includegraphics[width=0.45\textwidth]{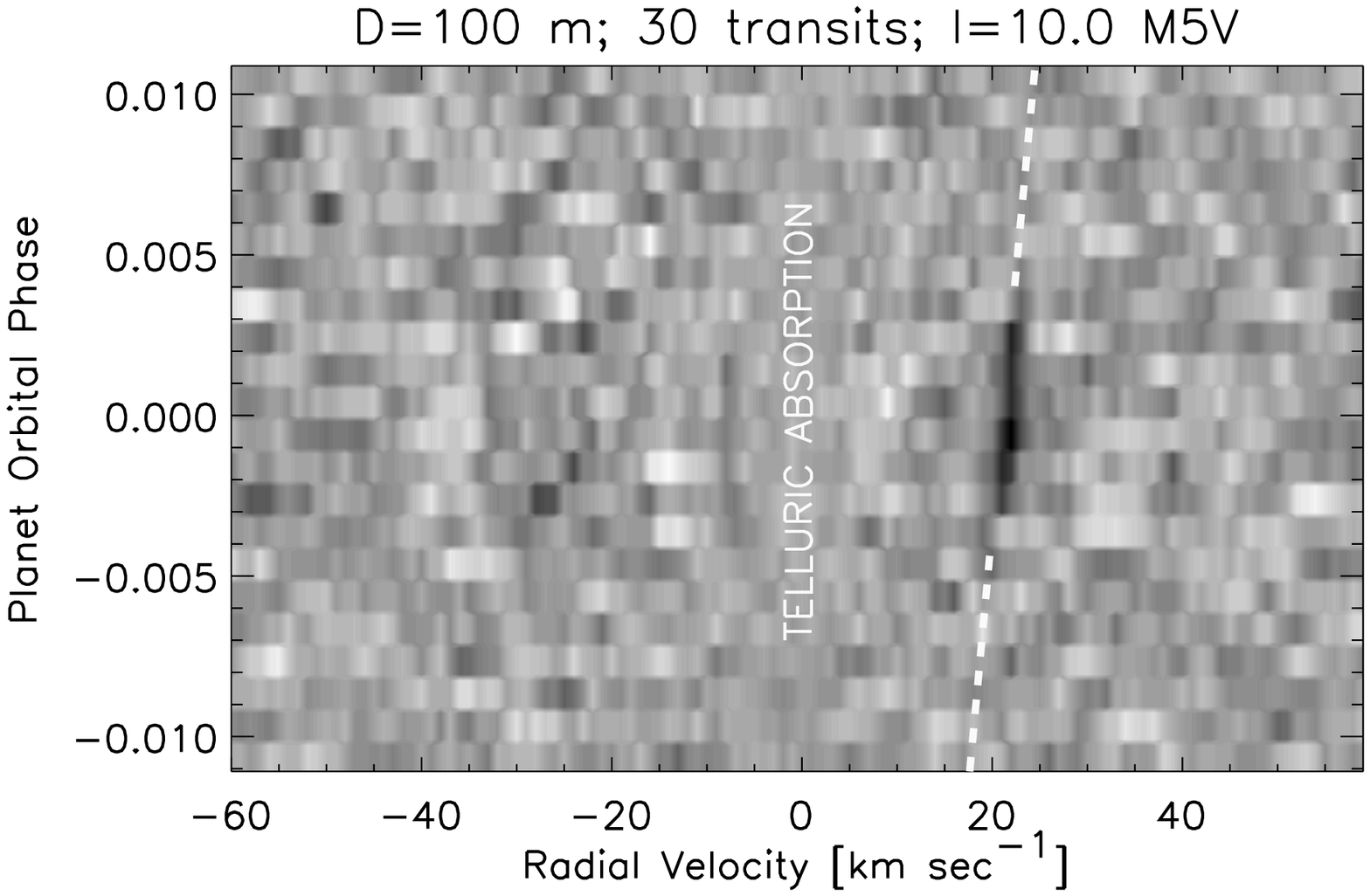}
\end{center}
\caption{\label{simo2} (left panel) The simulated O$_2$ signal from a hypothetical Earth-twin in the GJ1214 system, showing a 3.8 $\sigma$ detection after collecting data for 30 transits with the 39m E-ELT. The spectrograph is assumed to have a spectral dispersion of R=100,000 with a total throughput of 20\%. The simulated data were treated in the same way as the existing data of HD209458 and $\tau$ Bo\"otis (Snellen et al. 2010; Brogi et al 2012), and the telluric and stellar components removed. The signals from all O$_2$ lines in each spectrum are combined through cross-correlation with the model spectrum, and the results of the spectra taken during transit were ultimately added together. The procedure was repeated to emulate the observations of multiple transits. The right panel shows the O$_2$ cross-correlation signal as function of orbital phase for 30 transit observations of a similar system orbiting an I=10.0 M5 dwarf, taken with (an array of) flux collectors with a combined collecting area equivalent to that of a telescope with a diameter of 100 m. The dark slightly tilted strip at $+$21 km sec$^{-1}$ is the O$_2$ transmission signal. The dashed white line indicates the radial velocity of the planet as function of time, well separated from the telluric O$_2$ absorption at V=0 km sec$^{-1}$.}
\end{figure}

We performed a detailed simulation of observations of an Earth-twin transiting a late M-dwarf, using the planned 39-meter European Extremely Large Telescope (E-ELT). As host star we chose the M5.5V dwarf star GJ 1214, which has a transiting warm super-Earth (Charbonneau et al. 2009) in a 1.5-day orbit. In this system, a planet with similar insolation as the Earth should be at an orbital distance of 0.054 AU, implying an orbital period of 11.8 days. The I=11.1 magnitude of the host-star (V-I = 3.6) puts it within the expected 1$\sigma$ magnitude range for the brightest M4-M6 systems for Earth-twin planet frequencies of $\eta_e$ $\ge$ 25\%. 

We simulated observations for a telescope with a 39m diameter aperture, using a spectrograph with a spectral resolution of R=100,000, assuming a total system through-put of 20\%. For the stellar spectrum we used the publicly available\footnote{Available at http://kurucz.harvard.edu} Kurucz high resolution spectrum of the M5V dwarf GL 411. Atmospheric transmission was taken from Wallace \& Livingston\footnote{Available at ftp://ftp.noao.edu/catalogs}. Observations of 6 hours in length were simulated centered on the 1.6 hour transit, using 360 exposures of 60 seconds, of which 97 during transit. To simulate the influence of the change in altitude of the target as closely as possible, the airmass was varied between 1 and 1.8, scaling the atmospheric transmission accordingly. The system velocity of the host star is +21.1 km sec$^{-1}$ (Berta et al. 2011), meaning that the barycentric velocity of the Earth can be near zero (around opposition) without the telluric absorption overlapping with that of the planet. We let the earth barycentric velocity change by 100 m sec$^{-1}$ per hour to mimic the Earth rotation. For other stars it may be necessary to observe significantly away from opposition such that the barycentric velocity of the earth helps to shift the planet signal away from that of our own atmosphere. 

One potential issue is that the change in the radial velocity component of the planet during transit is significantly smaller, in this case from -0.9 km sec$^{-1}$ to +0.9 km sec$^{-1}$, than for the hot Jupiter HD209458b ($\pm$15 km sec$^{-1}$). Observing at higher spectral dispersion than R=100,000 will allow the same data analysis techniques to be used as by Snellen et al. (2010) and Brogi et al. (2012). Fortunately, the transit of a twin-Earth around a mid-M dwarf is relatively short, meaning that the baseline outside the transit can be used (when there is no planet signal present) to effectively remove the telluric absorption without making use of the change in the radial velocity of the target during transit. Note that the noise contribution from star spots is expected to be very small, because the technique is only sensitive to relative high-frequency changes in the spectra, i.e. changes in the stellar line shapes. Although M-dwarfs can vary in flux by a few percent on the time scale of weeks, this corresponds to absolute flux variations at the 10$^{-4}$ level during transit, and relative changes in the line depths at a significantly lower level.  Furthermore, the stellar spectra do not contain molecular oxygen lines meaning that any stellar line variability  will not add-up constructively during cross-correlation. On the other hand, stellar flares are abrupt in nature, and would make the data unusable.

The stellar model spectrum is scaled such that it corresponds to that of an I=11.1 star, and subsequently converted to units of photons per pixel. The Earth transmission spectrum as calculated above is scaled according to the radius of the star, and subsequently blue/red shifted and multiplied to each spectrum. This product is subsequently multiplied with the telluric spectrum appropriately scaled by airmass. Gaussian noise is subsequently added to each pixel value, scaled to the square root of the photon count to emulate Poisson noise. The resulting data are subsequently treated as the existing data of HD209458b and tau Bo\"otis b, and the telluric and stellar components removed. The signal of each spectrum taken during transit is subsequently added together to obtain the final result. The procedure was repeated many times to emulate observations of multiple transits. 

The left panel  of Figure 3 shows the simulated O$_2$ signal from a hypothetical Earth-twin in the GJ 1214 system, after collecting data for 30 transits with the E-ELT. Oxygen is detected at 3.8$\sigma$. From one particular site, typically three transits per year will be visible at favorably low airmass, meaning that an oxygen detection for this hypothetical planet could be reached in about a decade. In the most favorable case, if a transiting Earth-twin would exist around an I=10 M5V host star, this will be reduced to 3-4 years, using a dozen observing nights. In contrast, it will take $>$100 years before data from enough transits can be combined to reach a similar detection of an Earth-twin transiting a sun-like star. 

We compared the ELT high-dispersion transmission spectroscopy method with possible observations of biomarker gases with the James Webb Space Telescope (JWST - expected to be launched in 2018), as discussed by several authors (e.g. Kaltenegger \& Traub 2009; Rauer et al. 2011; Belu et al. 2011; Deming et al. 2009). The JWST can only perform relatively low-resolution (R$<$2700) spectroscopy at these wavelengths, decreasing the change in the effective planet radius due to O$_2$ transmission to 5 km (taking into account clouds). Compared to the O$_2$ signal at high dispersion, a detection will require approximately 10$^3$ more photons but over a 10$^2$ times wider wavelength range. Taking into account the smaller collecting area of the JWST and the likely higher throughput of the low-dispersion spectrograph, E-ELT transmission spectroscopy will be two orders of magnitude faster than low-dispersion spectroscopy with the JWST targeting the O$_2$ absorption band in the same hypothetical planet atmosphere. It should be noted that since the Earth-atmosphere does not interfere, it is approximately 10x more efficient for the JWST to target instead the 9.6 $\mu$m O$_3$ during secondary eclipse.  Overall this makes the E-ELT about an order of magnitude more efficient than the JWST in detecting biomarker gases in Earth-twin atmospheres.

\section{Flux Collector Telescopes}

While the E-ELT and the other planned extremely large telescopes will be built to provide high sensitivity and the highest possible angular resolution and image quality over a significant field of view, in principle only the collecting area aspect is relevant for high-dispersion spectroscopy of bright stars. In addition, the observational technique used is only dependent on the relative strength of narrow spectral lines, putting no requirements on absolute spectral photometry or on low-order spectral stability (Snellen et al. 2010; Brogi et al. 2012). This opens up the possibility to design (arrays of) relatively low-cost telescope systems, so called flux collectors, with poor image quality but very large collecting areas. A collecting area equivalent to two hectares would be sufficient to detect the example planet system discussed in the main text during one single transit.

The use of flux collectors for optical astronomy is still in its infancy. For our purpose the image quality must be such that most starlight is focused into an area of less than a hundred arcseconds squared. Even during full moon, the contribution of the sky in a 5-arcsecond diameter aperture is less than 3\% of that from an I=12 magnitude star, having little or no contribution to the overall noise budget. Telescope systems with relatively poor image quality are for example Cherenkov telescopes like the High Energy Stereoscopic System (H.E.S.S.) which consists of four 12 meter and one 28 meter telescope (Cornils et al. 2003) with point spread functions (PSF) in the order of ~100 arcseconds, which is insufficient. Sub-mm telescopes, if they would have an optical reflective coating, have a better image quality. The primary dish of the 15m James Clerk Maxwell Telescope has a surface accuracy of 20-30 micron RMS (Hills et al. 1985), resulting in a PSF of a few arcseconds. During the design phase it was once considered to give it a reflective surface so it could also serve as an optical telescope and rival the power of the 5m Yale telescope (R. Le Poole, private communications). We foresee no great engineering challenges in building arrays of large flux collectors with PSFs of a few arcseconds and a collecting area significantly larger than the ELT, for a fraction of the cost. For illustration, the right panel of Figure 3 shows the oxygen cross-correlation signal as function of orbital phase, assuming a 100 m diameter telescope system combining 30 transits for an I=10.0 M5 dwarf. 

It will require a significant development in spectrometer design to enable efficient high-dispersion spectroscopy for flux collectors. For conventional Echelle spectroscopy the size of the grating is proportional to both the size of the primary mirror of the telescope and the angular size of the entrance slit. This makes Echelle spectrographs already very large for the future ELTs, but would make them huge for flux collectors with a $>$1 arcsecond PSF. In the design of ELT Echelle spectrographs this is partly mitigated through the use of adaptive optics, which significantly decreases the PSF of the telescope, but this is by definition not applicable to flux collectors. If Echelle spectrographs are to be used, extreme image slicing techniques are needed to reduce the size of the spectrometer. Further size reduction could be achieved using immersed gratings such as planned for the METIS instrument (Lenzen et al. 2010) on the E-ELT. 

Fourier Transform Interferometers (FTS) form a possible alternative. A classical FTS is a Michelson interferometer scanning a movable mirror over some distance, measuring the Fourier transform of the spectrum. The major advantage of an FTS over an Echelle spectrometer is that its size does not need to scale with the telescope collecting area, allowing very compact designs even for ELTs or larger telescopes. A significant disadvantage is that for shot-noise limited observations, the noise will be proportional to the square-root of the power in the total observed wavelength range. In the case of absorption spectra, this means that the resulting signal-to-noise is a factor $\sqrt{m}$ smaller than for Echelle spectrographs, where $m$ is the number of sample points, if all other factors are equal. In the case of high-dispersion O$_2$ transmission observations m=4$\times 10^{3}$, meaning that a classical FTS is more than an order of magnitude less sensitive than Echelle spectrographs, removing all the benefits of using a telescope with a larger collecting area. In principle, this multiplex disadvantage can be largely mitigated by incorporating a dispersive element in the FTS design. In such a design the angular information is traded in to spread the interferogram frequency components over the detector such that significantly fewer sample points are combined per interferogram (Douglas 1997; Edelstein \& Miller 1999). This has the additional advantage that the spectrograph has no longer moving elements and that the spectrum is measured instantaneous at a sensitivity potentially comparable to Echelle spectrographs (Edelstein \& Miller 1999). 

 It may be beneficial to go to even higher spectral resolutions than R=100,000. This is the case if the planet O2 lines are unresolved at such dispersion. In general, the line-width is governed by the equatorial rotation velocity of the planet, which would allow some room to increase the spectral dispersion. However, in case of the Earth, the depth of the high-resolution transit signal is limited by a strong decrease in the abundance of molecular oxygen at the top of the mesosphere at 85 km, meaning there is no gain in increasing the resolution beyond R=100,000. However, this is only for exact copies of the Earth. We can only speculate about how the oxygen abundance varies as function of altitude on other life-bearing planets (e.g. Rauer et al. 2011). In addition, rotational velocities may be an order of magnitude lower for rocky planets in the habitable zones of red dwarfs stars, since they could well be tidally locked (e.g. Tarter et al. 2007). 
 
\section{Discussion}

It is a very long, difficult, and costly process to prove new technological concepts for highly innovative space missions such as Darwin and TPF (e.g. Beichman et al. 2006; Traub et al. 2006; Cockell et al. 2009). In addition, the field of extrasolar planets is continuously undergoing rapid developments, making it extremely perilous to forecast the state of play during operations of such missions many years in advance. In this paper we show that ground-based telescopes provide an interesting alternative for finding biomarkers in exoplanet atmospheres through high-dispersion transmission spectroscopy. If Earth-like planets are common, future ELTs can detect oxygen within a few dozen transits. In addition, we argue that in the future, dedicated flux-collector telescopes can be significantly more powerful than the ELTs for this particular science case, and be built for a fraction of the cost. These can ultimately provide the large collecting area needed to perform statistical studies of life-bearing planets in the solar neighborhood. 

It is yet largely unknown what fraction of M-dwarfs host Earth-like planets in their habitable zones (Bonfils et al. 2012), and it will require significant efforts to find the transiting planet population around the nearest and brightest stars, both from space and the ground (e.g. using PLATO - Rauer \& Catala 2011; TESS - Ricker et al. 2010; MEarth - Charbonneau et al. (2009); NGTS \footnote{http://www.ngtransits.org/}; MASCARA - Snellen et al. 2012). A clear advantage of ground-based high-dispersion spectroscopy over the Darwin and TPF-type space missions is that over the next decade the former technique can be further utilized, tested, and improved without expensive and high-risk investments, and can naturally develop into larger and more complex instrumentation over time. Each improvement in sensitivity will allow new types of observations to be conducted at the forefront of exoplanet research. Current instrumentation can probe, next to carbon monoxide, water and methane absorption in the infrared,  and possibly metal oxides, metal-hydrates, and reflected starlight in the optical wavelength regime. Furthermore, molecular line profiles can potentially show the effects of a planet's rotational velocity. 

Both optical and infrared Echelle spectrographs will become available on the ELTs during the first half of the next decade.  From the three planned ELTs, the Giant Magellan Telescope (GMT) has chosen a high resolution spectrograph, G-CLEF (Szentgyorgyi et al. 2012) as a first light instrument. The European ELT has the METIS (Brandl et al. 2010) and optical HIRES\footnote{http://www.eso.org/sci/facilities/eelt/instrumentation/index.html} instruments as third and possibly fourth light instruments. The combination of these telescopes and instruments will typically be an order of magnitude more powerful than those currently used. This means that the studies performed today on a handful of planets can be extended to a few hundred hot Jupiters. The phase function of the brightest targets can be studied, which is directly linked to their global atmospheric circulation, revealing changes between a planet's morning and evening spectrum driven by photochemical processes. Furthermore, atmospheric studies can be extended to temperate and smaller planets. We believe that before the ELTs come online, a purpose-built 5-10m flux collector telescope can on the one hand deliver competitive science as a dedicated instrument for exoplanet characterization. On the other hand it could serve as a proof of concept and test-bed for flux collector technology.  

Note that ultimately the presence of oxygen in a rocky planet atmosphere in itself will not be sufficient to claim existence of extraterrestrial life. For example, it has been claimed that, under certain circumstances significant amounts of oxygen can be produced by abiotic photochemical processes as well (Leger et al. 2011). Detection of oxygen should be seen as a first step, and further evidence for biological processes will require a detailed investigation of the planet atmospheric composition as a whole, and the use of possible further statistical arguments on the abundance of planets with oxygen-rich atmospheres as function of planet-mass, level of stellar insolation and stellar activity. Although the ground-based studies proposed here will only be sensitive to Earth-twins around M-dwarfs, these stellar hosts make up 80\% of the main sequence stars in the solar neighborhood. Since we can only speculate about how life developed on Earth, we dismiss the hypotheses that biological activity cannot occur on planets orbiting M-dwarfs as premature (Segura et al. 2010; Tarter et al. 2007)  - we just do not know.  The discovery of rocky planets in the habitable zones of their M-dwarf host stars with atmospheres showing strong oxygen absorption features will pave the way for detailed further studies with other means, including those of Earth-twin planets orbiting solar-type stars. The mapping of all chemical processes in these planet atmospheres, including targeting a range of other biomarkers, will ultimately be required to prove the existence of extraterrestrial life.

\acknowledgements

We thank the anonymous referee for helpful comments on the manuscript. I. S. acknowledges support by VICI grant no. 639.043.107 from the Netherlands Organisation for Scientific Research (NWO).

\end{document}